# Software Engineering Meets Systems Engineering: Conceptual Modeling Applied to Engineering Operations

Sabah Al-Fedaghi and Mahdi Modhaffar
Computer Engineering Department, Kuwait University, Kuwait

**Summary**
Models are fundamentally crucial to many scientific fields, including software engineering, systems engineering, enterprise modeling, and business modeling. This paper focuses on diagrammatic conceptual modeling, as opposed to mathematical or computational models, wherein a conceptual model is a translation of reality processes into an abstract mechanism that has similar structure and parallel events of the external processes. Although various modeling approaches exist, including UML (Unified Modeling Language) in software engineering and its dialect, SysML (System Modeling Language), in systems engineering, several difficulties arise in such models, including the problem of model multiplicity that is related to the lack an integrated view of structure and behavior. This paper generalizes conceptual modeling to be applied in organizations at large. According to authorities, the so-called organization theory portrays organizations as machine-like systems. As a machine, an organization coordinates its parts to transform inputs into outputs. Therefore, we synthesize the notion of an organization as a machine and apply a new modeling methodology called *thinging machine* (TM) to real engineering operations. The results show the viability of the TM methodology serving as a foundation for high-level modelling of systems.

## 1. Introduction

Modeling in systems and software engineering [1][2][3] concerns with the process of building a representation of the designated part of the world being investigated (domain). This paper focuses on diagrammatic conceptual modeling, as opposed to mathematical or computational model, wherein a conceptual model is an abstract representation of a system intended to replicate a part of a system and its behavior. Various approaches have been used for conceptual modeling, including UML (Unified Modeling Language) [4] in software engineering and its version, SysML (System Modeling Language), in systems engineering [5]. In this paper, we adopt a new modeling language, *thinging machine* (TM). TM modeling was originally proposed for use in the requirement analysis phase of the software development life cycle (SDLC) in software engineering. In this paper, we apply the TM methodology to model operations in system engineering, thus suggesting TM can play dual modeling roles, as shown in Fig. 1, with an implicit attempt to recommend

- Using TM notions to complement current models (UML, SysML, and OPM; e.g., adopting generic TM actions and viewing time/events at a higher level of specification)
- Using TM as a unifying (across many fields of study) modeling language.

We generalize conceptual modeling for application to organizations at large. According to common understanding [6] [7], *organization* is a generic term that exists in numerous forms, such as a business or a governmental department, enterprise/company, firm, etc. An organization has a structure, functions, processes, workers, and a purpose. The general theme adopted in this paper is the notion that an organization is a machine.

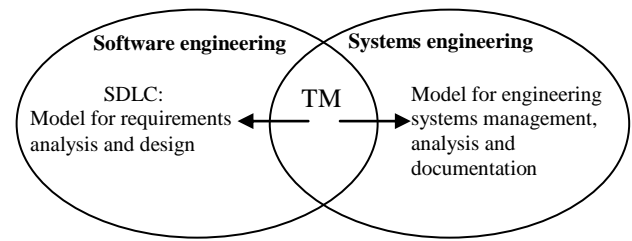

Fig. 1 TM modeling can be applied to real engineering processes and can be used as a first phase of developing information systems.

### 1.1 Problem: Difficulties with Standard UML/SysML

According to Brooks [8], "The hard part of constructing software [is] the specification, design, and testing of [the] conceptual model, not the work of representing it and testing the consistency of the representation." The inherent difficulty of the software development process is a reason for many software failures [9]. In this sense, a major issue arises during the requirements process of the software development lifecycle. "We are also having difficulty getting the specifications right," Marasco [10] states, "We don't do a good job of describing what we want created. A lot of the time, requirements are vague, undefined, incomplete, or contradictory. And developers often make educated guesses about what is desired, only to have to go back later in the cycle and rework their program due to discrepancies."

Oliver [11] claims, "UML is a powerful and versatile modeling language that can be used to model, not only object-oriented applications, but also application structure, actions, and business processes." Based on Oliver [11], there is *no holistic or adequate replacement* for UML. According to Evermann [12], "UML is ideal for conceptual modeling, "But the modeler must take particular care not to confuse software aspects with aspects of the real world to be modelled." UML has increased in sophistication, leading many people to believe they will be better off without it [11]. "Our coworkers don't see the difficulty that they have developed, and no amount of experience or expertise can slay critical complexity," states Fairbanks [13]. According to Dori [14], as the inherent complexity and interdisciplinary nature of systems increases, the need for a universal modeling, engineering, and lifecycle support approach becomes ever more essential. The unnecessary complexity and software orientation of UML calls for a simpler, formal, generic paradigm for systems.

The sheer number of UML diagram forms remains a source of concern. The problem of model multiplicity [14] concerns the integrated view of structure and action involved in the way UML diagrams are linked to one another. According to Soffer et al. [15], UML lacks a system-theoretical ontological basis encompassing assumptions about common features characterizing structures regardless of domain because it developed bottom-up from object-oriented programming principles. Moreover, the Object-Process Language (OPM) [16] was introduced as a formal and intuitive modeling language in this context. OPM is a holistic approach to systems engineering that unifies function, structure, and behavior in a single model. ISO/PAS 19450 [17] is a specification for OPM, which can be used instead of UML.

SysML [18], a dialect of UML 2, is based on UML. We will discuss UML and SysML as a single approach to conceptual modeling. SysML addresses systems engineering needs and is more suitable "to analyze, specify, design, and verify complex systems … to enhance systems quality, improve the ability to exchange systems engineering information amongst tools, and [to] help bridge the semantic gap between systems, software, and other engineering disciplines" [19]. SysML is utilized in model-driven systems engineering to analyze, specify, design, and verify complex systems with the aim of bridging the semantic gap between systems, software, and other engineering disciplines. However, SysML results in a "fragmented process" with hybrid diagrammatic descriptions and notations (e.g., blocks, activities, and uses) with no overarching design that brings the components together [20]. According to Mark Simons, Arena Technologies (https://www.arenatechnologies.com/is-sysml-right-for-se/), "OMG updated the UML metamodel for SysML and the result was a very verbose language, familiar to software engineers but too often rejected by other engineering disciplines, and completely incomprehensible to the average system stakeholder."

As suggested in this paper, UML and SysML methodologies succeed in modeling projects because of the variety of their representations, which are heterogeneous forms of diagrammatic notions. Yet, they fail to provide a nucleus from which various phases of the engineering process develop. Regardless of how important various perspectives on the framework are, a fundamental need still exists for an overarching paradigm that connects views and events into a unified, vertically multilevel conceptual structure.

### 1.2 Aim: A Conceptual Blueprint for Systems/Software Engineering

Abstraction is an important tool in the design and implementation of complex systems, but the process of abstracting excludes certain details to concentrate more on important aspects in the domain. One of the most essential techniques of computing is abstraction [21], which is required for the design and implementation of complex systems [22].

In the fields of systems and software engineering, the current trend is to create models with higher levels of abstraction. This involves developing a framework that represents a real organization's structure and operations. Such a conceptual model excludes technological aspects such as hardware and software. It acts as a blueprint for software/systems engineering that serves

- The requirements analysis processes of the software system in progress. In requirements engineering, building such a model is considered a bridge to the design and construction phases.
- A basic frame for identifying business processes and the way these processes are interconnected to achieve the organization's final objectives. The scheme is created alongside the daily operations of the business process. It provides process visualization and documentation to assist in defining work patterns, avoiding redundancy, or even designing new processes.

### 1.3 Generalizing the Perspective of Research: Organization as a Machine

This research broadened software/systems modeling to include all types of organizations. "Organization is everything, and everything is organization," some claim in various contexts [23]. The term *organization* comes from the word *organism*, which refers to a body structure divided into pieces that are kept together as one organic whole by a web of relationships [24]. To function, an organization communicates with and adapts to its environment in the same way a biological organism does. According to Herbert Simon, an organization's primary tasks are making decisions and "making things happen" [25]. "Consider an organization to be a collection of restrictions on the activities undertaken by a group of coordinating agents," as Fox et al. put it [26]. From the ontological perspective, organizations are differentiated into basic configurations and they involve many structures such as aims, job processes, authority, roles, and communication [27].

An organization is made up of subsystems that communicate with one another to operate appropriately. To ensure sustainability, an enterprise strives to consider and influence the external world while aligning its internal subsystems to meet the demands of the environment. The division of labor, according to Luther [28], is the cornerstone of any company. A shoe factory, for instance, might have workers employed in divisions such as leather cutting, eyelet stamping, tops stitching, soles sewing, feet nailing, lace adding, and shoe packing; the factory's role as an organization is to connecting the various work divisions together. Organization as a means of coordination necessitates the establishment of authority with a goal of enterprise that is realized by the collective efforts of many people in a specific time and place.

The latest contributions to the organization theory, according to Starbuck [29], have depicted organizations as machine-like structures. An organization, like a computer, coordinates its parts to translate inputs from an outside world into outputs for that outside world. Such adoption of the computer metaphor paradigm is a recent conceptual advance. Vuorinen et al. [30] coined the "as-a-machine" principle to create an information security theory that provides a theoretical foundation for understanding the way an object can be protected: "The machine metaphor offers a broad perspective on information security in which no one has absolute power; it is always somewhere in the middle." The use of a computer as a symbol is a well-known phenomenon. Additionally, living objects are also compared to complex machines. In other words, they have properties that are not found in inanimate matter. In demonstration, according to Dalio [31], the economy is a computer.

In this context, *enterprise engineering* has arisen as a new discipline of engineering that looks at businesses from an engineering point of view. Specifically, enterprise architecture has its roots in both organizational and information technology sciences [32]. Vries et al. [33] describe the field of engineering as a discipline that addresses enterprise design holistically. The term *enterprise* refers to a type of business organization [6]. According to Aveiro [32], "The operating theory of companies is that participants enter into and comply with agreements, and in doing so bring about the business services of the enterprise." Enterprise engineering, therefore, is becoming a discipline that addresses enterprise architecture holistically [33].

**1.4 Contribution: New Conceptual Modeling Approach for Organizations**

This paper synthesizes the notion of an organization as a machine and the new TM conceptual modeling methodology as an integrated blueprint for software, systems, and organization engineering (see Fig. 2). This notion is applied to two operations of an oil tanker company (OTC). The OTC's fleet-engineering process provides the oil tankers with various types of services in two oil ports. The activities of this process are determined by the number of tankers calling at two oil ports, which in turn is dictated by the volume of export and types of vessels. Multiple physical operations are performed in the fleet-engineering process, one of which is highlighted and adopted as Case Study 1 (vessel berthing process) in Section 3. A second case study involves a cargo oil filling process discussed in Section 4.

Of course, due to space and time limitations, the research in this paper presents parts of the OTC's processes and organization. The aim is to show the viability of the proposed integration approach that involves systems, software, and organizational aspects.

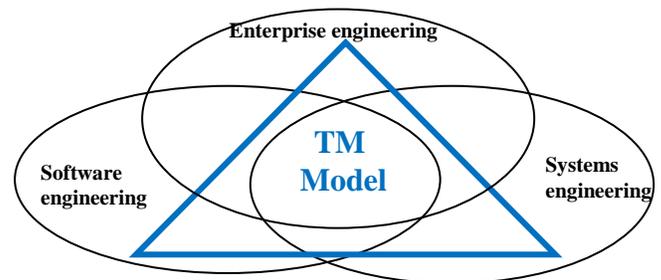

Fig. 2 TM as an integrated model in software engineering, systems engineering, and enterprise engineering.

**1.5 Overview**

The remainder of the paper is outlined as follows: In Section 2, TM is described and its terminology and importance to modeling systems are discussed in detail. Furthermore, two case studies will be presented in Sections 3 and 4. Each case study is described in detail, including their static descriptions, dynamic specifications, and behavioral models.

## 2. Thinging Machine Modeling

Currently, a prominent ontology (e.g., object-oriented) views the world as static entities with spatial and causal relations with one another. On the other hand, in a processual ontology (e.g., Whiteheadian), the world is conceptualized as dynamically stabilized processes. In this paper, the proposed thing/machine ontology views the world from both perspectives. Static things with defined assemblies (machines) exist in terms of five potential actions they employ in interaction (flow) with other things. Thus, "being" is built from the potential five actions. Entities (called *thimacs*; i.e., thing–machines) persist by means of embedding time to form higher-level entities (events). Entities embedded with time are events.

Such a thing/machines ontology is built upon the observation that any human encounter with the world needs a "model" as a mediator (see source in [34]). According to Albert Einstein [35], "It seems that the human mind has first to construct forms independently, before we can find them in things." Thus, a model can only be invented or derived from other inventions. Modeling may involve setting of non-fully observable structures and such structures cannot be fully "discovered": They can only be invented [34].

The TM model (see [36] and its sources) started with exploration of research into types of information and reached the conclusion that information can only be created, processed (changed), released, transferred, and received. Generalizing such a thesis, the TM model is a conceptualization of the ways things can be merged into an assembly of interrelated things. Accordingly, the TM ontology (*ontology* here refers to an explicit specification of a conceptualization) is built from fundamental entities called *thimacs* as a dualistic form of a *thing* and a *machine*. A machine is an arena of creation, processing, releasing, transferring, and/or receiving things. The arena is a thing that can be an object to machines. The arenas form a field (specified as a TM diagram) that includes fully observable structures "space/time presence" and non-fully observable structures and it subsumes both "existing" things (objects) and occurring events. A thimac extends across (occupies) space-like region (static TM model) while progressing over time (TM events). This is close to Whitehead's process philosophy that considers an event a form of apprehending being. A thimac that is realized as an event (called *process*) is a TM thimac with a time subthimac (see source in [34]).

Crossover between thingness and machinery occurs when the thimac role changes between being a flow-actor (create, process, release, transfer and/or receive) and being an object of flow (being created, processed, released, transferred and/or received). All things (thimacs) are created, processed, and transported (acted on), and all machines (thimacs) create, process, release, transfer and/or receive other things (i.e., thimacs; see Fig. 3). Create, process, release, transfer and receive are called (potential) *actions* or *stages* in a machine. Things "live" or "pass through" other machines. Machines house other things and provide roads for their flow. The unity of thing and machine forms a thimac. In such a blend, a single thimac is a fusion of two manifestations. The thing flows within machines, and it serves as a machine for other flowing things. The machine in Fig. 3 is more complete than the known input-process-output model is.

As shown in Fig. 3, a TM machine can be viewed as a coordinated system of flow (a change in the action position). The flow is different from movement (e.g., in space) because the latter is relative to some arbitrary, "fixed" reference point, but things that exist in time flow by their very nature as an existence of what happens in time. According to van Fraassen [37], "What happens in time [dynamic], and what exists in time [ontological]; … these two ways of being in time are different."

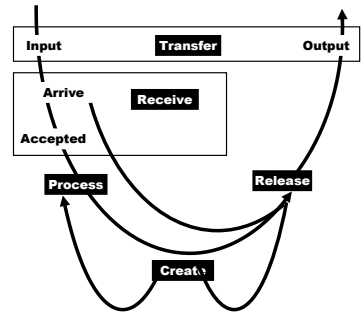

Fig. 3 Thinging machine.

Thimac trajectories of flow (all possible paths) include all available time "locations" (extensions in the five actions) while the thing actually exists in the TM model. A thimac therefore has a defined "location" in its static TM diagram, but it is common to all the "available" paths of its flow. A machine has its things inside its five phases (actions). Each thing inside the machine has an entry (to the machine) stage (create or input) and a departure (destroyed or output) stage. Note that a certain stage of a machine may have several things stored simultaneously.

Fig. 3 can be described in terms of the following generic (i.e., cannot be reduced any further) actions:
**Arrive:** A thing moves to a machine.
**Accept:** A thing enters the machine. For simplification, we assume that all arriving things are accepted; hence, we can combine the arrive and accept stages into one stage: the **receive** stage.
**Release:** A thing is ready for transfer outside the machine.
**Process:** A thing is changed, but no new thing results.
**Create:** A new thing is "born" (being found/manifested) in the machine. Things come into being in the model by "being found." Creation in metaphysics involves bringing the entities from the state of nonbeing into existence. The TM model limits this creation to the appearance in the model.
**Transfer**: A thing is input into or output from a machine.

Additionally, the TM model includes the mechanism of triggering (denoted by a dashed arrow in this study's figures), which initiates a flow from one machine to another. Multiple machines can interact with each other through the movement of things or through triggering. Triggering is a transformation from one series of movements to another.

Conceptualizing a thimac as a thing presents no indication as to the content of the thing, whereas conceptualizing a thimac as a machine forces a definite structure of actions with flow of other things. The totality of the universe (of modeling) is also a thimac. In TM, a set and the relationships of objects within the set are also thimacs. For notational convenience, they can be drawn differently.

An important distinction in TM is between static thimacs and events. We would expect that the static description, as an organizational (structural)/formational/topographical level, does not specify the instances or events. In the static TM diagram/subdiagram (which will be called a TM *form*: a generic term for a static TM diagram), everything is there; nothing corresponds to time (past, present, or future), and nothing corresponds to, say, the principle of no contradiction. However, what is "there" is loaded with potentiality that can be exemplified by actuality. A TM diagram (which may be called a TM *dimension*) encompasses the material space and nonmaterial 'space.' Accordingly, TM flow is more general than physical movement is. A static thimac with a single (potential) action (i.e., create, process, release, transfer, or receive) can be called a *purely static* (time-stationary) thimac.

An event is a form (i.e., static thimac) that has a time "breath" that infuses the dynamism of the time thimac in the thimac form (subdiagram). Dynamism is an unfolding mechanism of the static form that aligns it with reality through such machinery as igniting and chronologizing actions, logicalizing, and executing/controlling processes. Dynamism involves the development of actuality and the realization of static form through time. Accordingly, a thing with a time subthimac is considered an instance (individual). Individuals are things that exist in space *and* time.

## 3. Case Study 1: Vessel Berthing

We apply TM modeling to real control-oriented processes in an OTC using the TM approach via two case studies: vessel berthing and cargo oil filling (COF). These physical processes involve vessel operations, including berthing, departure, maintenance, cleaning, and system analysis.

### 3.1 General Description of Vessel Berthing

From an organizational perspective, the operation and control of vessel berthing involves the following machinery:
- **Marine agency:** Generally, an oil-tanker-related marine agency is in charge of disseminating critical information to all parties involved in marine transportation, including the ship owners, charterers, port authorities, and customs agents. In the case study provided in the paper, the approaching vessel requests permission through a designated marine agency that diverts it to the head office.
- **OTC Head Office**: The OTC headquarters is in charge of ensuring the OTC's overall performance. The head office includes OTC executives who make the company's business decisions. They rely on the OTC information system (IT), provided by the OTC data center, for information in its decision process.
- **Data Center:** The data center is where the OTC's shared IT data resources and operations are located, and it is centralized for the purposes of storing, processing, and disseminating information.
- **Port:** A port facility consists of multiple areas where ships can dock to load and unload cargo and passengers in a berth area. Ports are responsible for a variety of environmental impacts on local ecologies and waterways, such as direct water-quality effects from dredging, spills, and other emissions.
- **Ship Tender:** The ship tender is a boat that is used to serve or assist ships, normally by carrying people or materials to and from shore or other ships. Tenders, also known as dinghies, are common on smaller vessels.

The berthing operation begins before the arrival of the vessel to the designated port, when the vessel sends an arrival email to the marine agency to obtain permission. After receiving a permission email, the vessel begins to approach the shore, awaiting a small boat coming from the ship's tender to guide it up. At the same time, an arrival notification email is forwarded from the marine agency to the head office's database to store it in the database.

Upon the arrival of the designated small boat, a berthing group is established by having the boat and the vessel move together towards the vessel's berthing spot, which lies in the berthing area of the shore. The vessel is maintained in the berthing area to ensure it is ready for the next journey.

After the completion of the maintenance process, the vessel managers inform the marine agency to obtain permission to depart on their next journey. Once the vessel manager receives the departure notification email from the marine agency, the vessel moves from its berthing spot and away from the shore. Simultaneously, the marine agency informs the head office of the departure procedure by forwarding the departure notification email to be stored in the OTC's database.

Fig. 4 shows the overall description of the COF process.

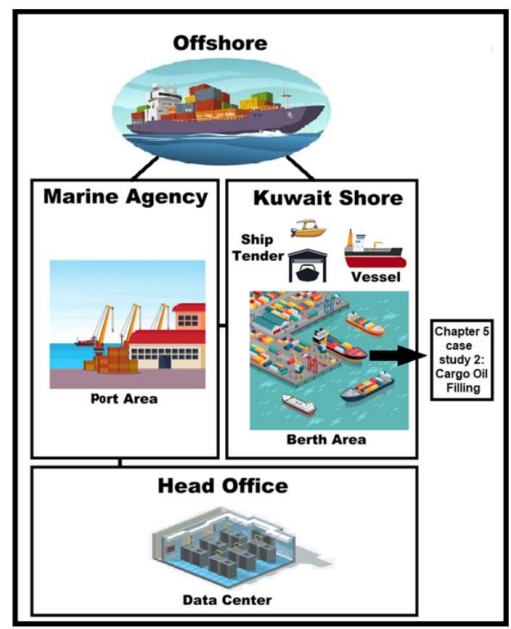

Fig. 4 The overall description of the COF processes.

## 3.2 Modeling Berthing

Currently, no explicit document contains a detailed procedure of the berthing operation. All vessel operations are usually passed verbally by superiors to individuals. TM modeling involves two levels, staticity and dynamics.

The static model involves spatiality and actionality (potentiality of actions). The dynamic level includes events and behavior.

### 3.2.1 The Static Model

Fig. 5 shows the TM static diagram of the berthing process.

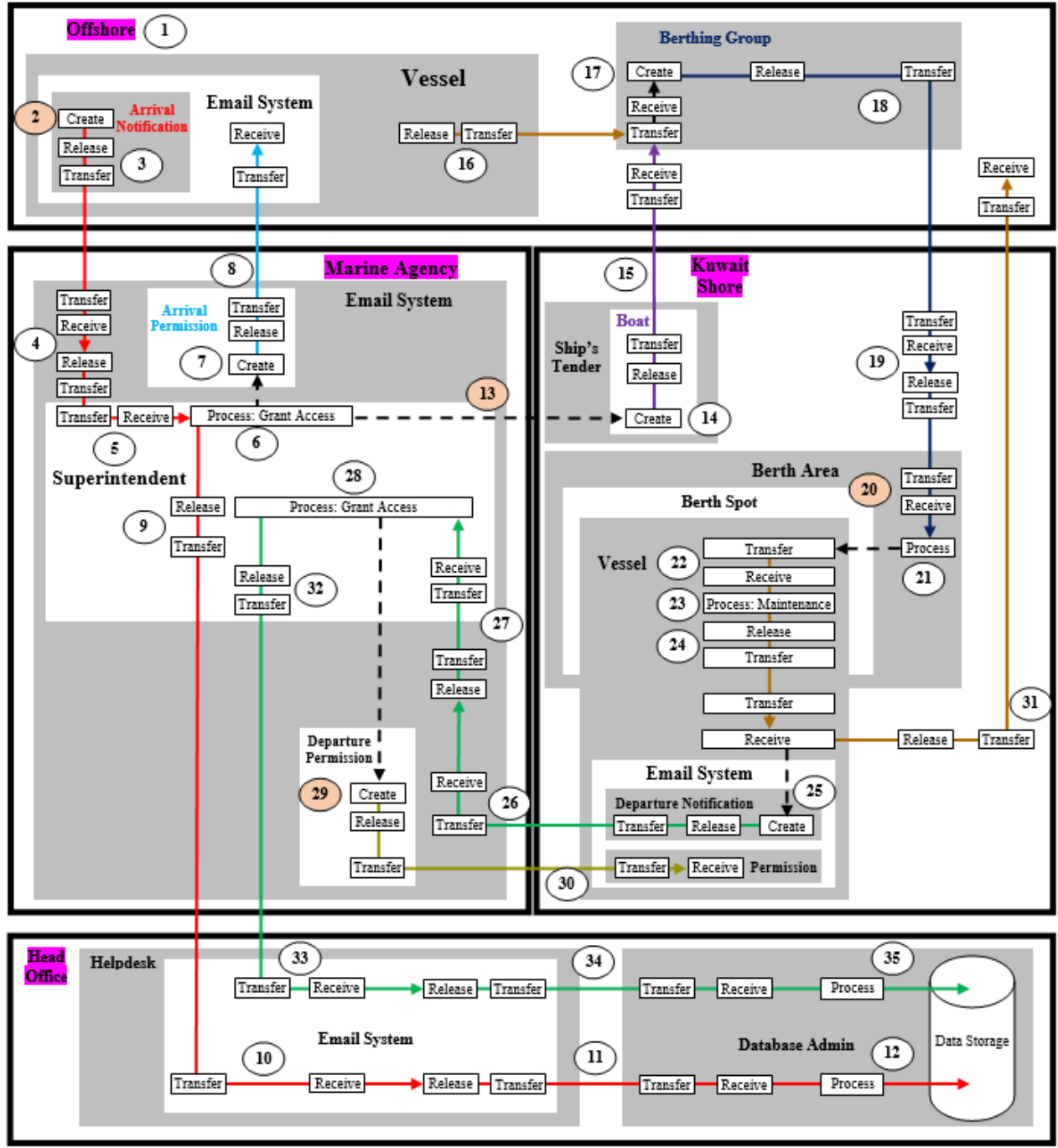

Fig. 5 The static TM model of the berthing process.

- In Fig. 5, a vessel, upon arrival to the offshore area (circle 1), creates an arrival notification email (2) and sends it to the marine agency's email system (3). The received email (4) is forwarded to the superintendent (5), where it is processed (6). Accordingly,
  - The superintendent creates an arrival permission email (7) and sends it to vessel (8).
  - The superintendent forwards the email to the head office (9), where it is received by the helpdesk's email system (10) that forwards it (11) to the database administrator (12) to be stored in the database.
  - The superintendent makes a call (13) to the ship's tender to prepare (14) and send a boat to the offshore area (15) to assist the involved vessel.
- The vessel moves to form a berthing group with the involved boat (16 and 17). The group, guided by the boat (18), moves towards the shore (19), and then into the berth area (20). The vessel moves to the berthing spot (21), where the vessel takes its spot (22).
- In the berthing spot, the vessel is processed as "in maintenance" (23). When the maintenance process concludes, the vessel leaves the berth area (24) into Kuwait's shore water.
- When the vessel departs, it sends a departure notification email (25) that flows (26) to the marine agency. The marine agency forwards the departure email to the superintendent (27), who begins processing (28) it. Accordingly,
  - The superintendent creates a departure permission email (29) and sends it to the vessel (30), which, in turn, departs Kuwait's shore towards sea (31).
  - The superintendent forwards the departure permission to the head office (32), where it is received by the helpdesk's email system (33). The system forwards the permission (34) to the database administrator to be stored in the database (35).

### 3.2.2 Dynamic Model

The decomposition of the static model lays the groundwork for specifying events. To construct the dynamic model, we identified the following events (refer to Fig. 6):

Event 1 ($E_1$): The vessel creates and sends an arrival notification email to the marine agency.
Event 2 ($E_2$): The marine agency forwards the arrival notification email to the superintendent.
Event 3 ($E_3$): The superintendent processes the arrival notification email.
Event 4 ($E_4$): The superintendent creates and sends an arrival permission email to the vessel.
Event 5 ($E_5$): The superintendent forwards the arrival notification email to the head office's helpdesk.
Event 6 ($E_6$): The helpdesk forwards the arrival notification email to the database administrator.
Event 7 ($E_7$): The database administrator processes and updates the database.
Event 8 ($E_8$): The ship's tender prepares a boat and sends it to the vessel location offshore.
Event 9 ($E_9$): The vessel moves to form a berthing group with the boat.
Event 10 ($E_{10}$): The berthing group moves to the Kuwaiti shore.
Event 11 ($E_{11}$): The berthing group arrives at the berthing area, where the vessel separates from the boat.
Event 12 ($E_{12}$): The vessel arrives at the berth.
Event 13 ($E_{13}$): Maintenance is performed on the vessel, and it is released back to the Kuwaiti shore.
Event 14 ($E_{14}$): The vessel creates and sends a departure notification email to the marine agency.
Event 15 ($E_{15}$): The marine agency forwards the departure notification email to the superintendent.
Event 16 ($E_{16}$): The superintendent processes the departure notification email.
Event 17 ($E_{17}$): The superintendent creates and sends a departure permission email to the vessel.
Event 18 ($E_{18}$): The vessel is released from the Kuwaiti shore into offshore waters.
Event 19 ($E_{19}$): The superintendent forwards the departure notification email to the head office's helpdesk.
Event 20 ($E_{20}$): The helpdesk forwards the departure notification email to the database admin.
Event 21 ($E_{21}$): The database admin processes and updates the database.

The behavioral model of the berthing system is shown in Fig. 7.

### 3.3 Analysis

The TM representation provides a conceptual basis for specifications and facilitates basic understanding of the berthing system's components. The model is an integrated static/dynamic description of the sub-systems of the berthing process that can be used to develop a control and management specification and analysis of engineering operation. The TM model can generally be seen as a way of "gaining control over the world processes" [38]. This includes understanding and assessing the current situation, diagnosing possible problems, designing changes, and realizing such changes, providing guidance for actors who operate in the involved system and expressing regulations [39].

The TM model can also be used to build an information system for the berthing system. The model presents a viable tool for creating a core scheme for describing requirements for these purposes.

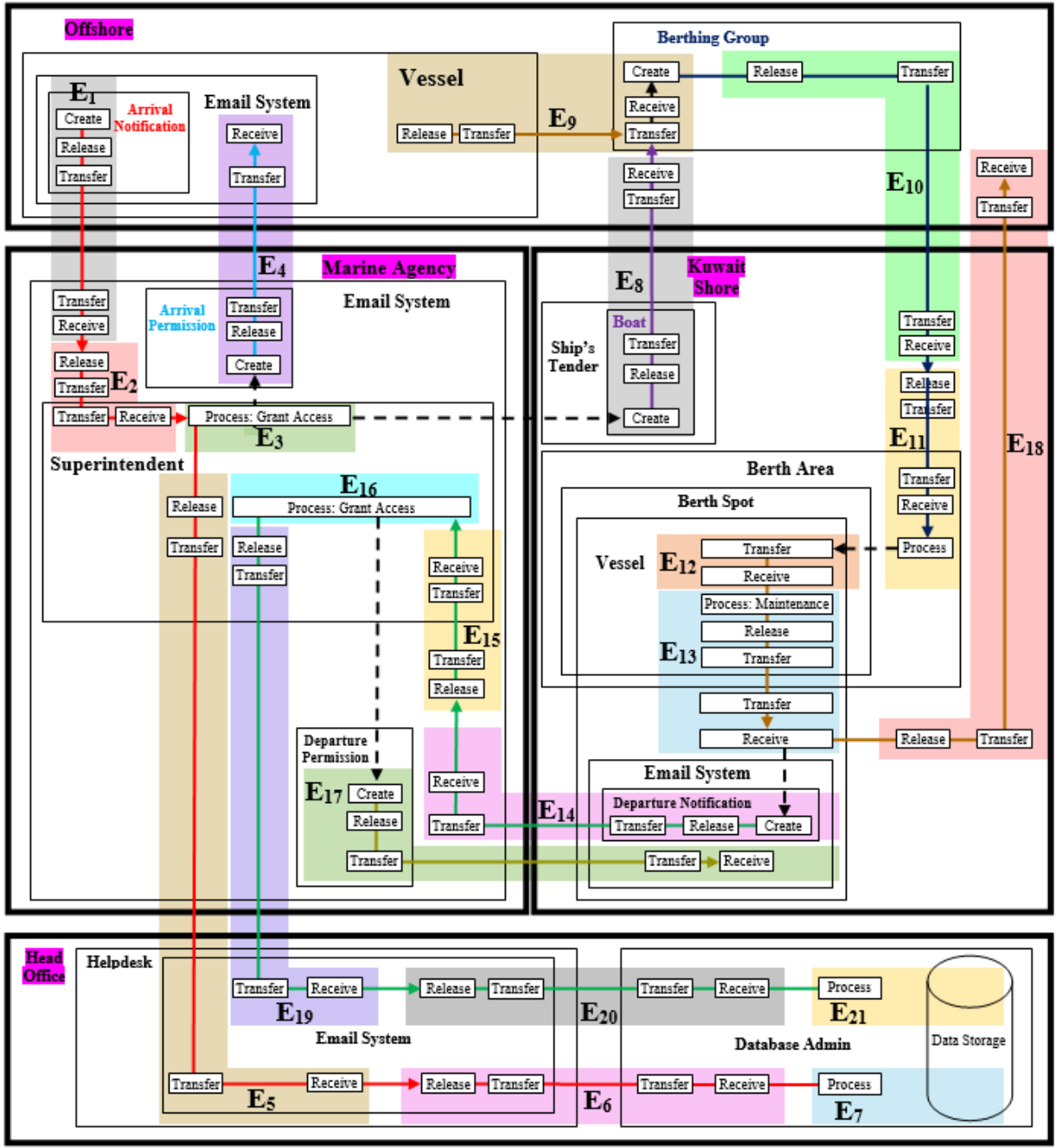

Fig. 6 Dynamic TM model of the berthing process.

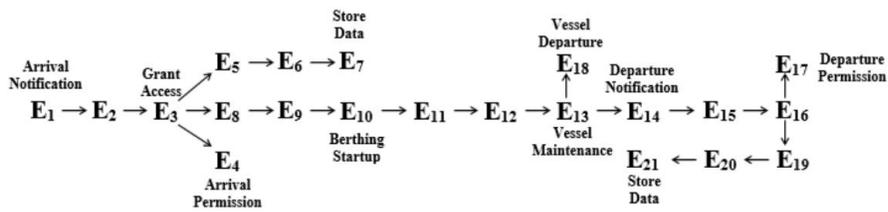

Fig. 7 Behavioral TM model of the berthing process.

## 4. Case Study 2: Cargo Oil Filling

This section presents a second case study that involves the most crucial activity in any oil transportation company, the COF operation. The main operator of any COF is the marine agency department. The physical operations executed by the marine agency involve: Vessel tracking, Vessel reporting process, COF, Cargo oil discharge, Robustness check and Full journey report.

### 4.1 General Description of Cargo Oil Filling

From an organizational view, the vessel COF process involves the following components and actors.

- **Oil Fields:** An OTC's oil field is a piece of land used to extract petroleum from the earth, such as natural gas or crude oil. Crude oil and petroleum products are often transported by tankers, pipelines, railway tank cars, and tank trucks, while storage is mostly done in surface or underground tanks.

- **Tanker Trucks:** An OTC's tanker trucks, also known as gas trucks or fuel trucks, are a type of truck used to transport liquids or gases on public roads. The majority of these vehicles resemble railroad tank cars, which are often used to transport liquids.

- **Port Oil Storage:** OTCs use oil storage for production, processing, and distribution, necessitating a wide range of storage tank types and sizes. Small bolted or welded tanks are suitable for manufacturing fields, whereas larger, welded storage tanks are commonly used in distribution terminals and refineries around Kuwait.

- **Surveyors:** Surveyors offer cargo safety advice and assistance, as well as ensuring that the oil meets contractual specifications. This process includes quality and quantity checks, storage facility inspections, conveying device inspections, and transport vehicle inspections, as well as supervising loading and discharge operations.

- **Oil Pumps:** Pumps and other equipment are used in the process of extracting oil and gas resources, refining them, and transporting them to where they are required. This holds true whether the oil and gas in question is sourced from traditional or unconventional reservoirs, onshore or offshore fields.

- **Loading Officers:** When a vessel is berthed in port, a loading officer acts as the marine transfer operator who supervises the transport of petroleum products among tanker ships, barges, and the terminal.

-

    During marine transfer activities, the loading officer ensures that all regulatory issues concerning environmental protection and maritime security are followed.

- **Vessels:** OTCs' vessels are designed to transport liquid cargo in bulk without the use of barrels or other containers inside their cargo spaces. The majority of OTC tankers transport crude oil to other ports worldwide that have oil-importing contracts with the OTCs.

### 4.2 Vessel Operation/Cargo Oil Filling Subsystem

In this section, we focus on the vessel operation/COF subsystem, which involves transporting crude oil from oil fields to fill vessels. The process includes many stages starting at the oil field, where the oil is extracted, stored in tanks, and prepared for the tanker trucks to pick it up. The OTC's marine agency makes sure that its portside oil tanks are filled and prepared for the arrival of any oil vessel to be filled for its next journey. The company's oil tanker trucks are filled from the oil field's tanks to transport oil to port.

When the tanker trucks arrive in port, they go to the oil storage area, where tanks wait to be filled. When filled, an oil sample is taken from the tanks and sent to the surveyor for processing. The surveyor's inspection process involves analyzing measurements such as density, temperature, and water dip readings; inspection concludes with an overall tank survey that is sent to the loading officer for assessment. Then, the loading officer creates a plan for cargo loading and safety procedures. In addition, the loading officer boards the visiting vessel and records vessel data, checks its systems, and creates a checking report. Subsequently, the loading officer takes his checking report portside, and a full loading plan is created and shared with the port crew.

Upon receiving a full loading plan, the technical crew moves to the equipment storage area, gathers any relevant hoses, and takes them to the oil pump area. Then, they take the first end of each hose and tighten it onto the oil dispenser valve. The other end is tightened onto the vessel's side, on an oil-tank-charging valve. After all preparations are completed, the crew creates a completion report and sends it to the loading officer, who gives the command to load. The loading process starts with the crew opening the oil storage tanks' discharge valves. This releases oil into the pump area pipes, then to the dispenser valve. Subsequently, the crew opens the dispenser valve to open a passage for the oil to flow through the hoses leading to the vessel's charging valves. These are opened to enable the vessel's oil tank to be filled for its next journey. Finally, after COF is completed, the crew creates a filling completion report and sends it to the port master.

## 4.3 Modeling Cargo Oil Filling

The COF operation is very complex and involves many operations working simultaneously and continuously to transport oil.

### 4.3.1 Static Model

Fig. 8 shows the TM static model of COF. It can be described as follows.

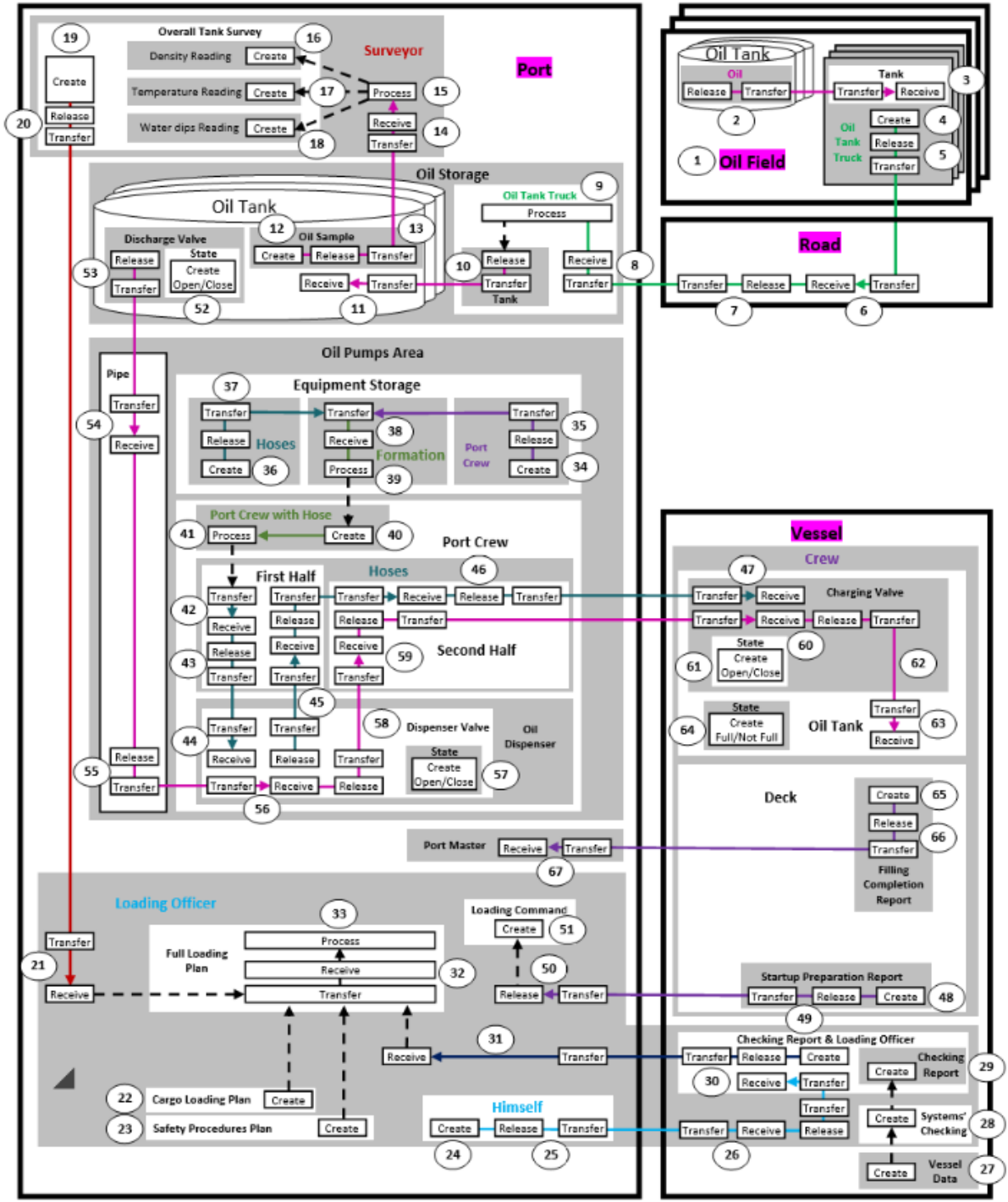

Fig. 8 Static TM model of the process.

*Transporting crude oil to port*
- In the oil field center (1 – upper right corner of the diagram), crude oil is stored in tanks (2) that receive oil via pipes from oil wells.
- The oil is released from the field's oil tanks (3) into trucks (4).
- Trucks are dispatched (4) from the oil field (5) to a road (6) leading to the port.
- The truck reaches the end of the road and arrives (7) at the OTC's oil storage area in port (8). There, truck's tank is processed (9) to transport (10) its oil to a local tank (11).

*Preparation for vessel filling*
- A sample of the oil is taken (12) and sent from the local oil tank (13) to the surveyor (14) for processing (15). Accordingly, the surveyor creates three types of data:
    - Density data (16)
    - Temperature data (17)
    - Water dip data (18)
- An overall tank survey report is created (19) by the surveyor after the data is taken. The report is sent (20) to the loading officer (21).

*Filling plan and checking the theater*
- The loading officer creates a cargo-loading plan (22) and a safety procedure plan (23) for the vessel filling process.
- The loading officer also boards the vessel (24) by going from the portside (25) to the vessel's deck (26).
- Once the loading officer is on deck, he receives and uses the vessel data (27) to check systems' functionality (28), then creates a checking report (29).
- Afterward, the loading officer takes the checking report (30) back portside (31), concluding the plan preparation (32), and processes it accordingly (33).

*Connecting the oil tank to the vessel*
- The crew starts moving (34) to do their job, by moving inside the storage room (35) and gathering hoses (36). These are taken (37) as a mobile hose unit (38) to be processed (39).
- The process includes moving the hoses into the oil pump area (40), where they are reprocessed (41) by extracting them from the mobile unit (42).
- One end of each hose is moved (43) next to the oil dispenser and equipped on its valve (44), while the other end of the hose is taken (45) as the other end of the hose (46) to be equipped on the vessel's charging valve (47).

*Assessing the connection operation*
- After the crew's job is completed on a vessel, a startup preparation report (48) is created and taken from the deck (49) to the loading officer portside (50).

- Upon receiving all necessary reports, the loading officer gives the command to load (51).

*Filling operation*
- Accordingly,
    - The crew converts the state of the oil tank discharge valve (in oil storage area) from off to on (52). This releases the oil (53) from the oil tank into pipes (54) in the oil pump area, and then from the pipes (55) directly to the oil dispenser (56).
    - The crew converts the state of the portside dispenser valve from off to on (57). This releases the oil (58) from the oil dispenser into a hose (59) connected to the vessel's charging valve (60).
    - The crew converts the state of the vessel's charging valve from off to on (61) and initiates oil tank filling by allowing the oil to move from the valve (62) into the tank (63).
    - Once the oil tank's state is converted to full (64), the crew starts (65) filling out a completion report and sends it (66) to the port master (67).

### 4.3.2 Dynamic Model

To construct a dynamic model, we identified the following events (see Fig. 9).

Event 1 ($E_1$): The oil is released from the oil field tanks into oil truck tanks.

Event 2 ($E_2$): The tanker trucks start traveling from the oil field to the road.

Event 3 ($E_3$): The oil tanker trucks reach the end of the road and enter the oil storage area at port.

Event 4 ($E_4$): The tanker trucks start the oil discharge process.

Event 5 ($E_5$): Oil is released from the trucks' tanks into the oil storage area tanks.

Event 6 ($E_6$): A sample is taken from the oil tank and sent to the surveyor.

Event 7 ($E_7$): The surveyor starts processing the oil sample.

Event 8 ($E_8$): Density, temperature, and water dip readings are taken to create an overall tank survey that is sent to the loading officer.

Event 9 ($E_9$): The loading officer creates a plan for cargo loading and safety procedures.

Event 10 ($E_{10}$): The loading officer boards the vessel.

Event 11 ($E_{11}$): The loading officer starts working on the vessel data, checking systems, and creating a report about them.

Event 12 ($E_{12}$): The loading officer completes the checking report.

Event 13 ($E_{13}$): The loading officer leaves the vessel with the report.

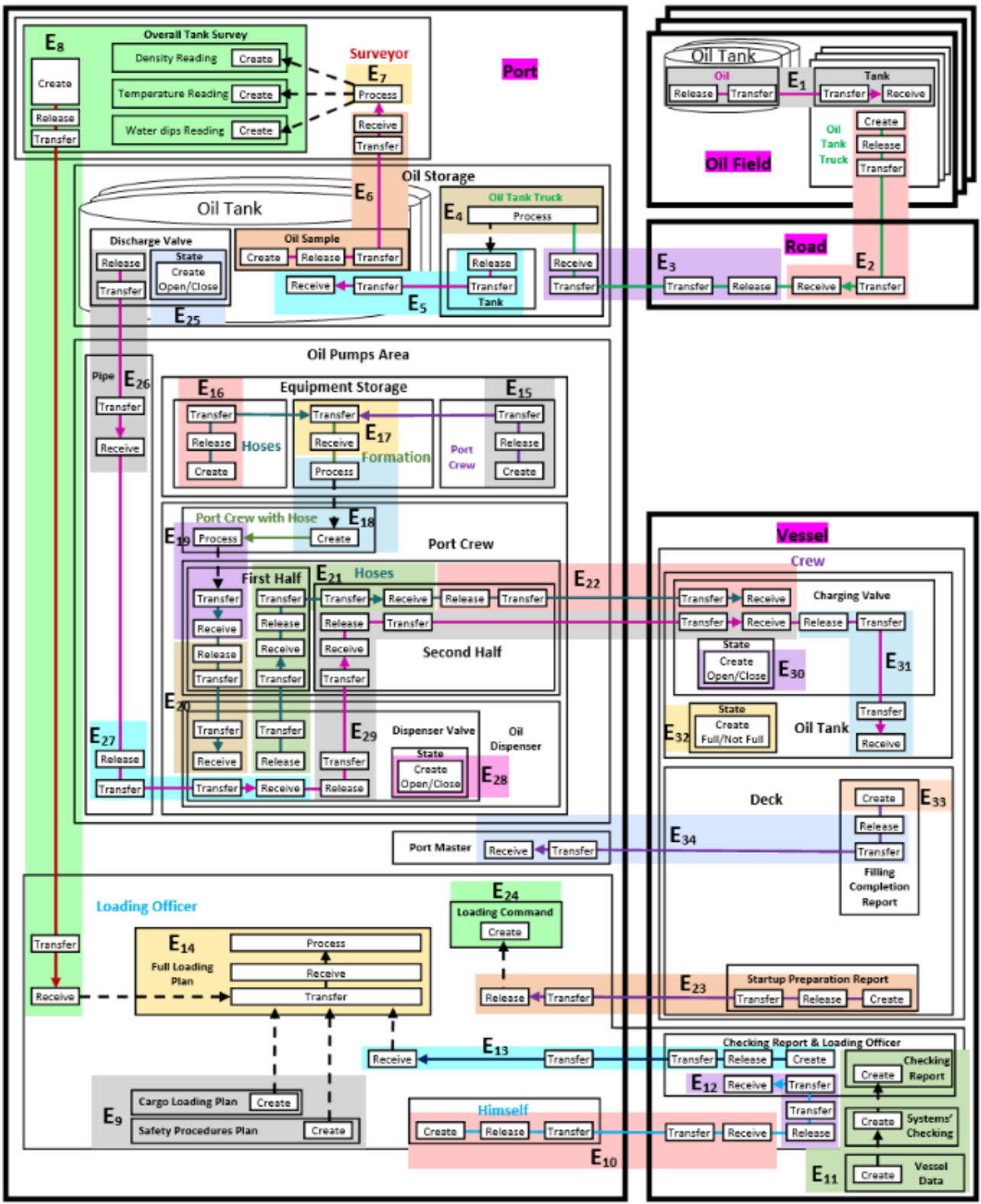

Fig. 9 Dynamic TM model of cargo oil filling process.

Event 14 ($E_{14}$): The loading officer creates a full loading plan and processes it by giving orders to the port crew.

Event 15 ($E_{15}$): The crew begins their work in the equipment storage area.

Event 16 ($E_{16}$): The crew takes hoses from the equipment storage area.

Event 17 ($E_{17}$): The "crew with hoses" formation is assembled.

Event 18 ($E_{18}$): The formation is processed by moving into the oil pump area.

Event 19 ($E_{19}$): The hoses are processed by the crew: the first end of the hoses is picked up.

Event 20 ($E_{20}$): The first end of the hose is taken and attached to the oil dispenser valve.

Event 21 ($E_{21}$): The other end of the hose is derived from the first half.

Event 22 ($E_{22}$): The free end of the hose is taken to the side of the vessel's hull and attached to its oil-tank-charging valve.

Event 23 ($E_{23}$): A startup preparation report is created and sent to the loading officer.

Event 24 ($E_{24}$): The loading officer gives the loading command to start the oil loading procedure.

Event 25 ($E_{25}$): The oil storage area tanks' discharge valves are opened.

Event 26 ($E_{26}$): Oil is released from the storage tanks into the pipes leading to the pump area.

Event 27 ($E_{27}$): Oil is released from the pipes into the oil dispenser valve.

Event 28 ($E_{28}$): The dispenser valve in the oil pump area is opened.

Event 29 ($E_{29}$): Oil moves through the hoses to the vessel's oil tank charging valve.

Event 30 ($E_{30}$): The charging valve is opened.

Event 31 ($E_{31}$): Oil moves into the vessel's tank.

Event 32 ($E_{32}$): The oil tank state is monitored until it is full.

Event 33 ($E_{33}$): A filling completion report is created by the crew.

Event 34 ($E_{34}$): The crew transfers the filling completion report to the port master.

The behavioral model of vessel operations/COF system is shown in Fig. 10.

### 4.4 Analysis

The resultant TM dynamic diagram can be used as a conceptual model in control, monitoring, and simulations. For example, monitoring can be applied to all events or subsets of them. Accordingly, when an event happens, it triggers a meta-event (an event that is caused by an event) to create a record that contains data about the time, changes in values, alerts, warnings, or any other needed information. The time data can contain various time information (e.g., start/end times, period). To use the data in a control scheme, a log manager can contain the full set of meta-events to create temporal log registration of historic archives of all events or to merge events into one bigger event.

Thus, the TM model matures vertically to represent time and to register changes in the system. The result is a clear systematic basis of a system's operations and alterations over time. Operations ensure [40] the implementation and control of activities and safe and reliable processes, as well as recognition of the status of all equipment and operators' knowledge and performance.

## 5. Conclusion and Future Work

TM is a modeling methodology with a simple ontology of five actions and two types of arrows applied uniformly to all stages of static, dynamic, and behavioral representations. This paper has demonstrated the applicability of TM modeling in an actual domain to model physical engineering systems and their behavior in an integrated way. TM results in a conceptual description that can be used for controlling the maintenance of a physical system. Additionally, TM can be used within a paradigm that views an organization as a machine, a concept borrowed from organization theory.

We claim the TM methodology can viably serve as a foundation for conceptual modelling of systems at large. Still, TM needs a great deal of effort to refine and apply it to more real-life systems. Further research should aim to develop complete TM specifications of such systems.

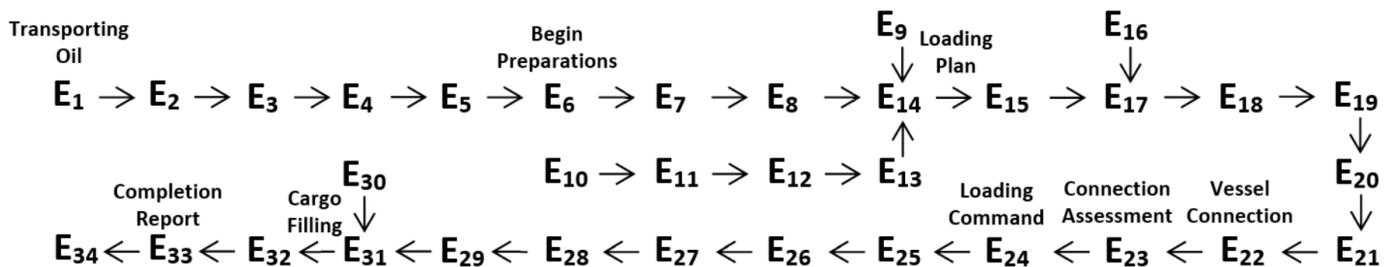

Fig. 10 Behavioral model of the cargo oil filling process.